# Elimination of Static Hazards in Asynchronous Sequential Circuits using Quantum dot Cellular Automata

Angshuman Khan[1, *], Chiradeep Mukherjee[2], Ankan Kumar Chakraborty[3], Ratna Chakrabarty[4], Debashis De[5]

[1, 2, 3] Department of Electronics and Communication Engineering, University of Engineering and Management, Jaipur, India
[4] Department of Electronics & Communication Engineering, Institute of Engineering & Management, Kolkata, India
[5] Department of Computer Science & Engineering, West Bengal University of Technology, Kolkata, India
[*] angshumankhan2910@gmail.com

**ABSTRACT**

There is nowhere else in emerging technology, but in Quantum-dot Cellular Automata, one can find high speed, low power operation, and high packaging density, which deals with electrostatic interaction between electrons within a cell. Literature survey lacks in hazards free design of QCA circuit. Hazards create ambiguous and unpredictable output, which can be avoided. This work considers both hazards and hazards-free asynchronous sequential circuits; both are compared in terms of kink energy, and a better one has been proposed. The circuit simulation has been verified in the QCADesigner tool.

## I. INTRODUCTION

Quantum dot Cellular Automata (QCA) is the most recent popular technology in the VLSI scale, which promises low power consumption, high packaging density, and very high speed for the design of logical circuits in Nano-level computing. The QCA was proposed by Lent &Tougaw, 1993 [1]. As semiconductor technology goes deeper into the 30nm lithographic process, the problem of leakage current and quantum effects becomes dominant in device physics. The zero-dimensional quantum cell with four dots in four corners mitigates these problems, employing quasi-adiabatic switching. The four electrons can take the farthest positions of a square to maintain highly stable configurations through Columbic interaction and quantum mechanical tunneling. The state-of-the-art power consumption forgets its way into QCA as electrostatic interaction requires no traditional current flow in the circuit. In this technology field, two excess electrons in either corner of a square provide highly stable configurations to encode logic 1 and 0 [2, 3, 4, 5]. The circuits implemented using this technology do not require traditional interconnections. Veterans have several times designed and tested the QCA circuits using molecular or magnetic QCA cells. Some researchers implemented molecular QCA structures which operated at normal temperatures [6]. The basic primitives of QCA technology are majority gate, inverter & wire. Different researchers have designed QCA circuits using majority gates [7]. The optimized QCA wire and inverter using a minimum number of cells are already proposed [8, 9]. The researchers also focused on the effect of temperature in multilevel QCA circuits [10]. Reconfigurable circuits have been designed in QCA technology [11, 12]. The QCA full adder using three inputs as well as five input majority gates was introduced [13]. Different memory circuits have been designed several times [14]. One of the most attractive topics in QCA is hazard-free circuit design. The constant output of a certain circuit instigates the condition for static hazard. A single transitory improper output due to the alteration in input literal when the output is likely to remain in the same state is called a static hazard. This paper discusses hazards in digital circuits first, then the effect of the same in QCA circuits. The hazard-free circuit is proposed and verified in the QCADesigner tool, which optimizes the stability of the circuit in terms of kink energy.

This paper has been arranged as follows: Section II introduces the fundamentals of QCA. The next section discusses hazards in Boolean circuits and gives an idea of how they can be optimized. Section IV sees the hazards in QCA circuits. Section V introduces the kink energy concept of QCA circuits. The simulation reports are in section VI. Finally, section VII concludes the work and discusses future work.

## II. FUNDAMENTAL OF QCA

A brief introduction to the background of QCA will help you understand this paper easily.

### A. QCA cell

The fundamental nanostructure of Quantum Dot Cellular Automata (QCA) is a cell. A QCA cell comprises four interacting quantum dots placed in the four corners of a square cell. Each of these dots is capacitively coupled together by means of a tunnel junction. The logical cell can contain two free electrons. When two excess electrons are charged, due to the Columbic force of repulsion, the two excess electrons occupy diagonal positions in a quantum Cell [15, 16]. The electrons always take the antipodal sites to be stable. Hence, the alignment of the electrons is of two types, and each of the alignments can be regarded as a polarization. The configuration of this electrical charge encodes binary information. Depending upon two types of polarization, the cell would have been viewed as steady state polarization of P= +1 and -1 [17-19]. There are two types of QCA cells, $90^0$ rotations and $45^0$ rotations, as shown in Fig. 1(a) and Fig. 1(b), respectively.

### B. Majority gate

The three-input majority gate is shown in Fig. 2. The logic function for the three-input majority gate is,

$$M(A, B, C) = AB + BC + CA \dots\dots\dots\dots\dots\dots\dots\dots(1)$$

The equation for AND gate is,

$$M(A,B,0) = AB \quad \dots\dots\dots\dots\dots\dots\dots\dots\dots\dots\dots\dots\dots(2)$$

The equation for OR gate is,

$$M(A,B,1) = A + B \quad \dots\dots\dots\dots\dots\dots\dots\dots\dots\dots\dots(3)$$

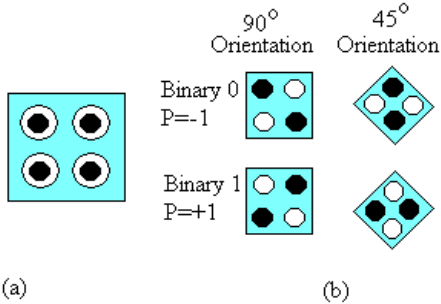

Fig. 1(a) QCA cell   Fig. 1(b) Different rotations of cells

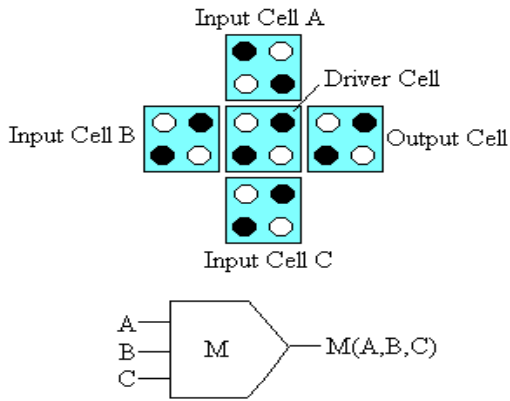

Fig. 2 Majority gate

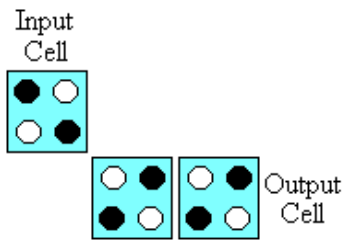

Fig. 3 QCA inverter according to [8]

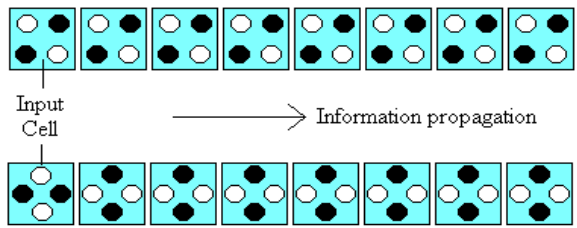

Fig. 4 Binary wire in different rotations

### C. Inverter

Fig. 3 shows the QCA inverter circuit, where the output returns the opposite polarity of the input. The simple, small, high-polarized inverter [8] is shown in Fig. 3.

### D. Wire

The QCA wire is shown in Fig. 4. In a QCA wire, binary signal propagates from input towards output due to electrostatic interactions between cells.

### E. Clocking

For proper functionality in a QCA circuit, a QCA clocking mechanism is required. The QCA clocking is easily understandable in the QCADesigner tool, where the clocking system is composed of four clocking zones, each synchronized with the others. Each clocking zone comprises QCA cells controlled by the same clock. The Fig. 5 shows the inter-dot barrier for a complete clock cycle. The four phases of a clock are the switch, hold, release, and relax phases; they have a rising edge, hold level, falling edge, and low level, respectively. In the switch phase, the input cells affected the un-polarized QCA cells, reaching a polarized value. In the hold phase, the polarization of the cell value is locked for transmission to the neighboring cells. During the release phase, QCA cells are losing their values. Finally, the relax phase considers the QCA cells to be completely unpolarized [1, 17].

### III. HAZARDS IN DIGITAL CIRCUITS AND ELIMINATION

In sequential circuits, the output depends on present and past inputs. When a clock does not control the circuit, the transition from one state to another occurs whenever there is a change in the inputs at any time; these circuits are called asynchronous sequential circuits. Unlike synchronous circuits, the state change of an asynchronous sequential circuit happens only when the input changes its state. It is customary to say that the clock signal in an asynchronous circuit is given to the first level of the circuit. For further clarification of the concept of static hazards in digital circuits, consider the arbitrary K map function,

$$f = A\bar{B} + B\bar{C} \quad \dots\dots\dots\dots\dots\dots\dots\dots\dots\dots\dots\dots(4)$$

For equation (4), the K-map will be like as shown in Fig. 6 below.

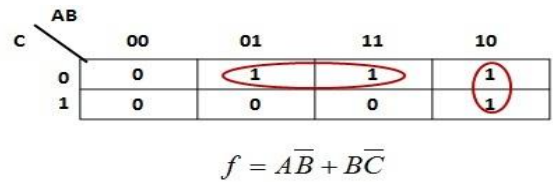

Fig. 6 K-map of $f = A\bar{B} + B\bar{C}$

According to equation (4), the corresponding digital circuit diagram is shown in Fig. 7. Here, the inputs are A, $\bar{B}$, and $\bar{C}$.

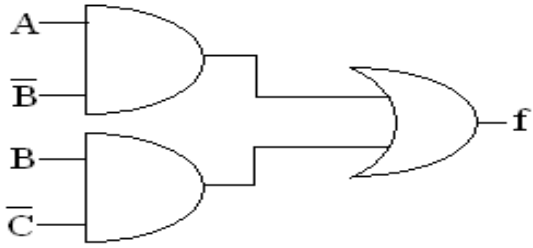

Fig. 7 Gate level design of $f = A\bar{B} + B\bar{C}$

The circuit has three inputs shown in Fig. 7: A, B, and C. The simplified circuit is shown in Fig. 8. It is the circuit with hazards. The reason for these hazards has been discussed in the next paragraph.

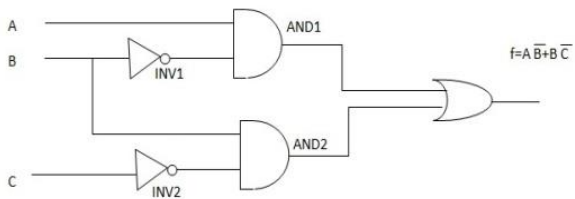

Fig. 8 Simplified circuit with hazards

The timing diagram of the circuit given in Fig. 8 is shown in Fig. 9. When the input 'ABC'= '100', the output, f = 1, due to the HIGH state at the output of the upper AND gate (AND1). For the input, 'ABC' = '110', the output *f* is HIGH due to the lower AND gate (AND2) output. So, in both cases, the output remains in a HIGH state or '1'. Due to the change in input 'B,' the HIGH output of the upper AND gate switched over to the lower AND gate. Hence, the output '*f*' is supposed to remain in the HIGH or '1' state. But due to unequal propagation delay of gates, if the upper AND gate output changes to '0' shortly before the lower AND gate output becomes '1', then during this short period, the output '*f*' will be '0'. This is the actual problem that occurs in circuits. It is called static Hazards [20, 21].

Fig. 7 shows digital circuits with hazards, but a hazard-free digital circuit is depicted in Fig. 8.

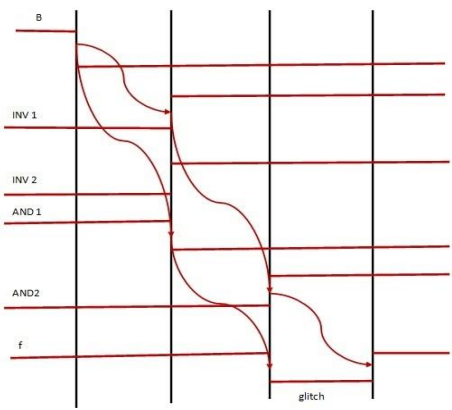

Fig. 9 Timing diagram

## A. Static Hazards Elimination

In the K-map, shown in Fig. 6, if the same product term does not cover the adjacent min-terms, then a hazard exists. The static hazards can be eliminated by grouping adjacent cells corresponding to 110 and 100, as shown in Fig. 10 below.

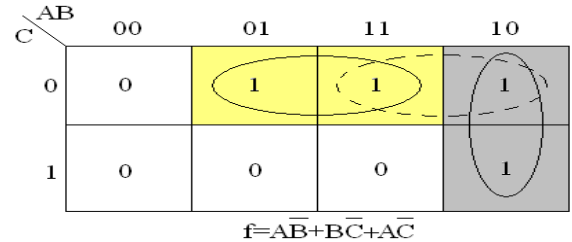

Fig. 10 K map for $f = A\bar{B} + B\bar{C}$ with adjacent cell groping

The extra product term does not include changing input variables, preventing possible momentary output glitches. The redundant term '$A\bar{C}$' has been added to the circuit to prevent hazards, as shown in Fig. 11.

When the input 'ABC' changes from '100' to '110', the output will remain in '1' or HIGH state because of the HIGH output at the new AND gate.

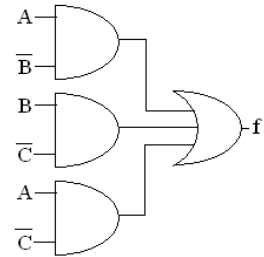

Fig. 11 Hazards free digital circuit for $f = A\bar{B} + B\bar{C} + A\bar{C}$

## IV. HAZARDS IN QCA CIRCUITS AND ELIMINATION

The hazard can pose difficulty in QCA circuits, which can affect in the same manner as in classical Boolean circuits. If the above digital circuits are designed in QCA technology, they face the same hazards problem. The QCA circuit for the K map in equation 4 is shown in Fig. 12, a QCA circuit with hazards.

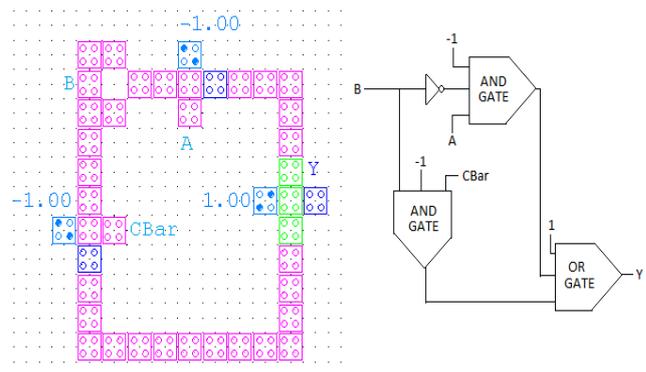

Fig. 12 QCA circuit with hazards

The above QCA circuit is designed in the QCADesigner tool. The logic operation of QCA circuits is the same as digital circuits (as discussed earlier), but for simplicity, the digital circuit is also shown beside the QCA circuit in Fig. 12.

### A. Hazards Elimination in QCA circuit

The procedure to eliminate the static hazards for digital circuits is discussed in this paper. The procedure to eliminate the hazards in QCA circuits is the same. The hazard-free QCA circuit, according to the K map shown in Fig 10, is designed in this paper. The hazard-free QCA circuit is shown in Fig. 13 below. To better understand the gate level equivalent, Fig. 13.

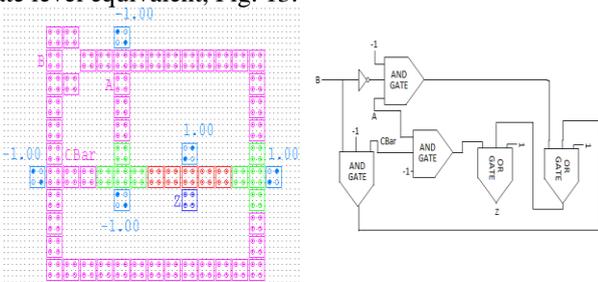

Fig. 13 Hazards free QCA circuit

## V. KINK ENERGY

The electrostatic interaction occurs between two QCA cells [8-11, 22], and it is given by

$$E = \frac{1}{4\pi\epsilon_0\epsilon_r} \cdot \frac{Q_1 Q_2}{r} = k \cdot \frac{Q_1 Q_2}{r} \quad\ldots\ldots\ldots(5)$$

Here, 'k' is constant, and its value is $9\times10^{-9}$. $Q_1$ and $Q_2$ are charges of electrons. So, the figure of 'E' becomes,

$$E = \frac{23.04\times10^{-29}}{r} J \quad\ldots\ldots\ldots(6)$$

This electrostatic interaction determines the Kink energy between two cells. Thus, Kink energy is defined as,

$$E_{kink} = E_{opp-polarization} - E_{same-polarization} \ldots\ldots(7)$$

The important point is that the kink energy between two QCA cells depends on the cell's dimensions and the spacing between the cells but not on the temperature [10].
Using the above equations, the kink energy of the outer portion of the Fig. 12 & Fig. 13 has been calculated here. For both cases, the kink energy equals $9.714 \times 10^{-20}$J. Fig 14 shows the output portion of the QCA circuits; whether it is hazard-free or not, the four cells under the round can take the total effect of the circuit. Hence, kink energy for those cells is calculated. The effect of the interaction of two electrons at the output cell, named 'x' and 'y', are shown in Fig. 15(a) and Fig. 15(b). The electrons e1, e2, e3, e4, e5, e6, e7 and e8 have different effect in the output cell.

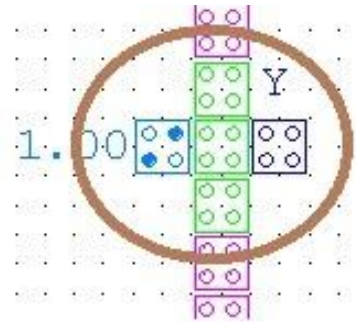

Fig. 14 Output portion of QCA circuit to calculate kink energy

$E_{opp-polarization} = 29.211\times10^{-20}$J
$E_{same-polarization} = 19.497\times10^{-20}$J
$E_{kink} = (29.211-19.497)\times10^{-20}$J $= 9.714\times10^{-20}$J

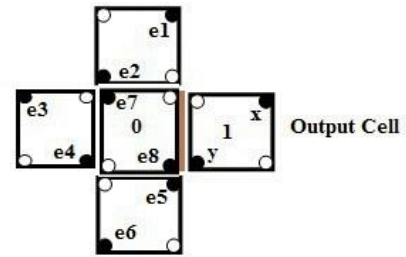

Fig. 15(a) Position of electrons at output portion for $E_{opp-polarization}$

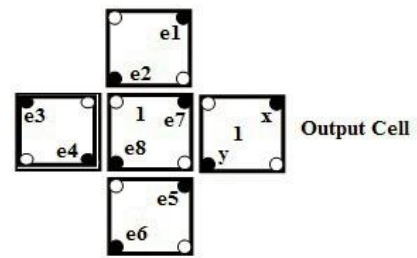

Fig. 15(b) Position of electrons at output portion for $E_{same-polarization}$

## VI. SIMULATION RESULTS

The proposed QCA circuits in this paper have been checked in the QCADesigner tool, version 2.0.3[23]. The default parameters for a bi-stable approximation are:

Number of samples: 12800
Convergence tolerance: 0.001000
Radius of effect: 65.000000 nm
Radius of permittivity: 12.900000
Clock high: 9.800000e-022 J
Clock low: 3.800000e-023 J
Clock amplitude factor: 2.000000
Clock shift: 0.00000e+000 J
Layer separation: 11.500000

143

Maximum iterations per sample: 100
Layer properties
Layer type: cells
Layer status: active
Layer description: main cell layer
Object properties: QCAD cell
Cell width: 18.00 nm
Cell height: 18.00 nm
Cell diameter: 5.000 nm
Default clock: clock0

Fig. 16 and Fig. 17 show the simulation results of the designed QCA circuit with hazard and without hazard, respectively.

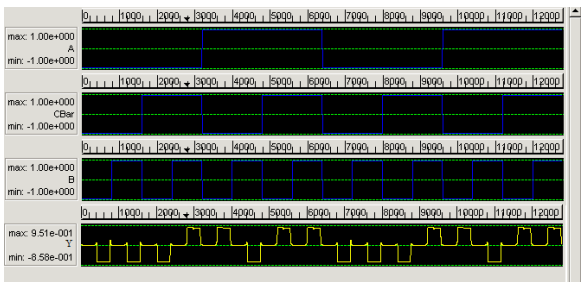

Fig. 16 Simulation results for QCA circuit with hazards

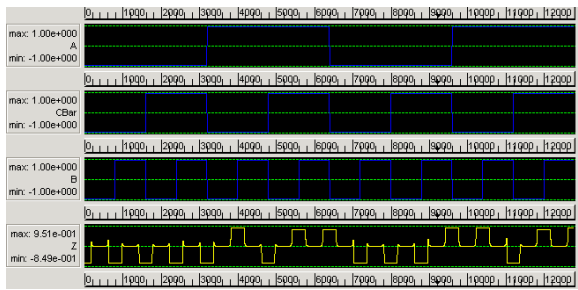

Fig. 17 Simulation results for QCA circuit without hazards

## VII. CONCLUSION AND FUTURE WORK

In this paper, the hazards in the QCA circuit have been discussed and designed in QCADesigner. This article also discusses the procedure to make the circuit hazard-free and the corresponding hazard-free QCA circuit. Depending upon the definition of hazards in QCA circuits, both hazards and hazards-free QCA circuits are implemented and verified at the same time with the same kink energy. Having the same kink energy, the hazard-free circuit is preferred so as to get better responses. So, this article proposes to consider a hazard-free circuit in QCA.

In the future, we can make different hazard-free QCA circuits. Not only sequential circuits but also the approach may be applicable for combinational circuits, irrespective of synchronous or asynchronous.

## ACKNOWLEDGMENT

The authors would like to express their gratitude to the anonymous reviewers.

## REFERENCES

[1] C. S. Lent, P. D. Tougaw, W. Porod, and G. H. Bernstein, "Quantum cellular automata," Nanotechnology, vol. 4, no. 1, pp. 49-57, 1993.

[2] C. S. Lent and P. D. Tougaw, "A device architecture for computing with quantum dots," Proceedings of the IEEE, vol. 85, no. 4, pp. 541-557, 1997.

[3] J. D. Wood and D. Tougaw, "Matrix multiplication using quantum-dot cellular automata to implement conventional microelectronics," IEEE Transactions on Nanotechnology, vol. 10, no. 5, pp. 1036-1042, 2011.

[4] M. Wilson, K. Kannagara, G. Smith, M. Simmons, and B. Raguse, "Nanotechnology: Basic science and emerging technologies," Champman & Hall/CRC, 2002.

[5] C. S. Lent, P. Tougaw, and W. Porod, "Bistable saturation in coupled quantum dots for quantum cellular automata," Applied Physics Letters, vol. 62, no. 7, pp. 714-716, 1993.

[6] W. Hu, K. Sarveswaran, M. Lieberman and G. H. Bernstein, "High-resolution electron beam lithography and DNA nano-patterning for molecular QCA," IEEE Transactions on Nanotechnology, vol. 4, no. 3, pp. 312-316, 2005.

[7] M. R. Bonyadi, S. M. R. Azghadi, N. M. Rad, K. Navi and E. Afjei, "Logic optimization for majority gate-based nanoelectronic circuits based on genetic algorithm," 2007 International Conference on Electrical Engineering, Lahore, Pakistan, 2007, pp. 1-5.

[8] R. Chakrabarty and A. Khan, "Design of a fault free inverter circuit using minimum number of cells & related kink energy calculation in quantum dot cellular automata," 2013 International Conference on Computation and Communication Advancement (IC3A), Kalyani, India, 2013, pp. 369-373, arXiv preprint arXiv:2310.10954 (2023).

[9] A. Khan and R. Chakrabarty, "Design of high polarized binary wires using minimum number of cells & related kink energy calculations in quantum dot cellular automata," International Journal of Electronics & Communication Technology, vol. 4, no. Spl. -2, pp. 54-57, 2013.

[10] R. Chakraborty, D. De, A. Khan, C. Mukherjee, and S. Pramanik, "Effect of temperature and kink energy in multilevel digital circuit using quantum dot cellular automata," 2012 5th International Conference on Computers and Devices for Communication (CODEC), Kolkata, India, 2012, pp. 1-4.

[11] A. Khan and R. Chakrabarty, "Design of Ring and Johnson counter in a single reconfigurable logic circuit in quantum dot cellular automata," International Journal of Computer Science and Technology, vol. 4, no. 1, pp. 363-367, 2013.

[12] R. Chakrabarty, D. Dey, K. K. Ghosh, "Design and analysis of Johnson counter using quantum dot cellular automata," 2011 International Conference at Institute of Engineering & Management, IEMCON 2011.

[13] K. Navi, R. Farazkish, S. Sayedsalehi and M. R. Azghadi, "A new quantum-dot cellular automata full-adder," Microelectronics Journal, vol. 41, no. 12, pp. 820-826, 2010.

[14] M. A. Amiri, M. Mahdavi, and S. Mirzakuchaki, "QCA implementation of a MUX-Based FPGA CLB," 2008 International Conference on Nanoscience and Nanotechnology, Melbourne, VIC, Australia, 2008, pp. 141-144.

[15] P. D. Tougaw and C. S. Lent, "Logical devices implemented using quantum cellular automata," Journal of Applied Physics, vol. 75, no. 3, pp. 1818-1825, 1994.

[16] A. O. Orlov et al., "Experimental demonstration of clocked single-electron switching in quantum-dot cellular automata." Applied Physics Letters, vol. 77, no. 2, pp. 295-297, 2000.

[17] S. Hashemi, M. R. Azghadi, A. Zakerolhosseini, and K. Navi "A novel FPGA-programmable switch matrix interconnection element in quantum-dot cellular automata," International Journal of Electronics, vol. 102, no. 4 pp. 703-724, 2015.

[18] V. C. Teja, S. Polisetti and S. Kasavajjala, "QCA based multiplexing of 16 arithmetic & logical subsystems-A paradigm for nano computing," 2008 3rd IEEE International Conference on Nano/Micro Engineered and Molecular Systems, Sanya, China, 2008, pp. 758-763.

[19] L. R. Hook, IV, and S. C. Lee, "Design and simulation of 2-D 2-Dot quantum-dot cellular automata logic," IEEE Transactions on Nanotechnology, vol. 10, no. 5, pp. 996-1003, 2011.

[20] G.K.Kharate, "Digital electronics," Oxford University Press 2010, New Delhi.

[21] S. Salivahanan, S. Arivazhagan, "Digital circuits and design," Vikas Publishing House Pvt. ltd., Noida.

[22] R. Farazkish, S. Sayedsalehi, and K. Navi, "Novel design for quantum dot cellular automata to obtain fault tolerant majority gate," Journal of Nanotechnology, vol. 12, 2012.

[23] QCADesigner, http://www.atips.ca/projects/qcadesigner, accessed online, January 2015.